\documentclass[conference,10pt]{IEEEtran}
\usepackage{fullpage}
\usepackage{enumerate} 
\usepackage{geometry}
\usepackage{graphicx,amsmath,amsfonts}
\usepackage[]{algorithm}
\usepackage{url}
\usepackage{algpseudocode}
\usepackage{float}
\usepackage{placeins}
\renewcommand{\baselinestretch}{1} 
\usepackage{array,booktabs}
\usepackage{graphicx,amsmath,amsfonts}
\usepackage{wrapfig}
\usepackage{geometry}
\usepackage{subfig} 
\usepackage{caption}
\usepackage{epstopdf}
\usepackage{cite}
\usepackage{setspace}
\usepackage{graphics}
\usepackage{amssymb}
\begin{document}
\title{Lossy Image Compression - A Frequent Sequence Mining perspective employing efficient Clustering}
\author{
\IEEEauthorblockN{Avinash Kadimisetty, C. Oswald, B. Sivaselvan}
\IEEEauthorblockA{Department of Computer Engineering\\ IIITDM Kancheepuram, Tamil Nadu, India\\
Email: \{coe12b009, coe13d003, sivaselvanb\}@iiitdm.ac.in}\\   
\and
\IEEEauthorblockN{Kadimisetty Alekhya}
\IEEEauthorblockA{Department of Computer Science \& Engineering\\ MVGRCE, Andhra Pradesh, India\\
Email: kalekhya366@gmail.com}}

\maketitle

\begin{abstract}
This work explores the scope of Frequent Sequence Mining in the domain of Lossy Image Compression. The proposed work is based on the idea of clustering pixels and using the cluster identifiers in the compression. The DCT phase in JPEG is replaced with a combination of closed frequent sequence mining and \textit{k}-means clustering to handle the redundant data effectively. This method focuses mainly on applying \textit{k}-means clustering in parallel to all blocks of each component of the image to reduce the compression time. Conventional GSP algorithm is refined to optimize the cardinality of patterns through a novel pruning strategy, thus achieving a good reduction in the code table size. Simulations of the proposed algorithm indicate significant gains in compression ratio and quality in relation to the existing alternatives. 
\end{abstract}
\section{Introduction}

Technological advancements led to the generation of large amount of data in the form of Text, Images, Audios, Videos etc. Huge amount of data is being transferred on the Internet everyday and the transfer time is highly affected by its size. Storing these large amounts of data is a challenging task and hence the need for compression arises. The number of bits required to represent the actual data will be reduced in the process of compression\cite{Salomon:2004:DCC}. Compression can be categorized into Lossless and Lossy. In Lossless compression, the actual data can be generated from the compressed data and the data is compressed by redundancy removal. Few examples include Huffman Coding, Arithmetic Encoding, Run Length Encoding, PNG, etc~\cite{huffman1952method, witten1987arithmetic, watson2003run, boutell1997png}. By contrast, in Lossy compression, only data approximate to the actual data can be constructed. It works by removing irrelevancy in addition to redundancy. Lossy compression achieves better compression than Lossless compression. Typical examples include MP3, MPEG, JPEG, JPEG 2000, codec2, Lossy Predictive Coding, GIF, etc \cite{Salomon:2004:DCC, Wallace1991, jpeg2000skodras}. Image compression is defined as the reduction in data required to represent the image and it is based on the fact that neighbouring pixels are highly correlated referred to as spatial redundancy. \\
Data Mining is the process of extracting hidden and useful information from large DB’s. Naren Ramakrishnan et al. observed five perspectives of Data Mining which are Compression, Search, Querying, Induction and Approximation\cite{RamakrishnanG99}. This paper focuses on the compression perspective of Data Mining. As the mined data can be viewed as a condensed representation of the original data, Data Mining can be viewed as a compression technique. Data Mining techniques include Association Rule Mining, Clustering, Classification, etc\cite{HanK2000}. In this paper, Clustering and Closed Frequent Sequence Mining are used to compress images.\\  
Clustering or Data Segmentation is used to partition large data sets into clusters based on their similarity \cite{HanK2000}. A cluster is the set of data objects similar to other objects within the cluster and dissimilar to objects in other clusters. Clustering algorithms can be categorized into Partitioning, Hierarchical, Density, Grid and Model based methods. The well known clustering algorithms include \textit{k}-means, \textit{k}-medoids, AGNES, DIANA, Birch, DBSCAN, etc. \cite{Jain2010kmeans,pamkaufman,HanK2000}. Frequent Sequence Mining is the process of mining complete set of sequences satisfying a minimum support threshold in a sequence database\cite{nbc2005}. Let $J = \{j_1, j_2, \dots, j_p\}$ be the set of items. An itemset is a subset of $J$. A sequence $\sigma$ = $\langle s_1, s_2, \dots, s_m \rangle$ ($s_i \subseteq J$) is an ordered list\cite{yan2003clospan}. The length of the sequence $l(\sigma)$ is the total number of itemsets in the sequence. If the length of the sequence is \textit{l}, it is called as an \textit{l-sequence}. A sequence $\sigma$ = $\langle x_1, x_2, x_3, \dots, x_m\rangle$ is a subsequence of another sequence $\theta$ = $\langle y_1, y_2, \dots, y_n\rangle$  denoted as $\sigma \sqsubseteq \theta$ iff $\exists k_1, k_2, \dots, k_m$ such that $1 \leq k_1 < k_2 < \dots <k_m \leq n$ and $x_1 \subseteq y_{k_1}$, $x_2 \subseteq y_{k_2}$,$\dots$,$x_m \subseteq y_{k_m}$. $\theta$ is called as \textit{super-sequence} of $\sigma$, and hence $\theta$ contains $\sigma$. In the context of this work, a sequence is considered to be ordered and continuous w.r.t to the index. Consider the text `\textit{encyclopaedia}'\ where `\textit{ecy}'\ is a sequence in conventional terms but in our work, `\textit{ency}'\ is only a sequence.
\section{Related Work}
Huffman coding and Arithmetic coding compress images by removing redundancy in coding of the pixels. A Huffman code is an optimum code where symbols that have a higher probability of occurrence have shorter codes than symbols with a low probability of occurrence\cite{huffman1952method}. Though simplicity and efficiency of Huffman coding makes it popular, it does not always produce the ideal codes because of its nature to assign codes of integral length to each symbol in the alphabet of the text to be encoded. This is overcome by Arithmetic coding technique where the entire input file is assigned a single code. In Arithmetic coding, initially each input file is represented as an interval of real numbers between 0 and 1 and the interval is narrowed on reading each symbol from the input file\cite{witten1987arithmetic}. Symbols with higher probability of occurrence narrow the interval lesser than the symbols with a lower probability of occurrence. A narrower interval requires more bits for representation and hence symbols with higher probability of occurrence are assigned shorter codes. Lempel-Ziv-Welch(LZW) algorithm is a dictionary based lossless compression technique in which the dictionary is initialized to all the symbols in the alphabet of the text to be encoded\cite{ziv1977universal}. LZW works by saving new phrases in empty dictionary entries and replacing phrases in the input text by pointers to the corresponding dictionary entries. 
\par Graphics Interchange Format(GIF) is developed as a modification of LZW encoding technique to compress image data. By nature, GIF is one-dimensional and it compresses two dimensional image data by working independently on each row of the image data. However, GIF is considered to be inefficient as it cannot find the relations between pixels of adjacent rows.  Transform coding is a technique for Lossy Image compression which works by dividing the original image data into smaller blocks. Each block undergoes a reversible transform process. This is followed by quantization of the transformed values and the blocks are encoded using Run Length Encoding, Huffman coding or other similar techniques. The most well known application of transform coding is JPEG(Joint Photographic Experts Group) standard for lossy image compression which uses the Discrete Cosine Transform(DCT)\cite{Wallace1991}. JPEG transforms color images from RGB to a luminance/chrominance color space and organizes the image into $8\times8$ blocks. DCT is applied to each block to obtain the frequency components followed by quantization using a quantization matrix. The result is encoded using a combination of Run Length Encoding and Huffman coding. The JPEG compression provides excellent compression rates at good quality. However, sometimes the reconstructed image of the JPEG algorithm has a blocky appearance. 

\begin{algorithm}[!htb]
\caption{\textbf{Compressor}}
 \label{algo-Compressor}
\begin{algorithmic}[1]
\State \textbf{Input:}{
	\begin{enumerate}
		\item The source image $I_{m \times n \times 3}$
		\item Block size $b$, Number of clusters $k$, Minimum Support $\psi$
	\end{enumerate}
}
\State \textbf{Output: }{$C_I$, the compressed encoded image}
\State Split $I$ into $I_1$, $I_2$, $I_3$ corresponding to Red, Green, Blue components respectively. 
\State Number of blocks $n_b$ = $\frac{m\times n}{b\times b}$
\State $i \gets 1$
\State{\textbf{parallel for }\textit{i} in \textbf{[1, 3]}}
\State \qquad Split $I_i$ into $n_b$ blocks $B_{i_1}$, $B_{i_2}$,$\dots$, $B_{i_{n_b}}$
\State \qquad $[T_{B_1},T_{B_2},\dots,T_{B_{n_b}}]$ = \textit{Parallel\_Clustering}($B_{i1}$, $B_{i2}$,$\dots$, $B_{in_b}$, $k$)
\State \qquad Replace each pixel value in every block with the cluster identifier of the cluster to which the pixel belongs.
\State \qquad $I_{i}^{'}$ = Merge all blocks $B_{i_1}$, $B_{i_2}$,$\dots$, $B_{i_{n_b}}$
\State \qquad $S_i$ = \textit{Modified\_GSP}($I_{i}^{'}$, $\psi$)
\State \qquad $S_{i}^{'}$ = \textit{Support\_Modified}($S_i$, $I_{i}^{'}$)
\State \qquad $CT_{i}$ = \textit{Huffman\_Coding}($S_{i}^{'}$)
\State \qquad For each sequence $\sigma$ in $S_{i}^{'}$, replace $\sigma$ in $I_{i}^{'}$ with its Huffman code to form $C_I$
\State \textbf{end for}
\end{algorithmic}
\end{algorithm}
\section{Proposed Architecture}
\subsection{Compressor}
\subsubsection{Source Image}
The approach of the proposed algorithm is mentioned in Algorithm~\ref{algo-Compressor}. $I$, the source image, is chosen in RGB color space and is converted to an $m\times n \times 3$ matrix. $I$ is split into three components Red($I_1$), Green($I_2$) and Blue($I_3$). The fact that each component is independent of each other is exploited in this paper. The algorithm is applied in parallel to each component to reduce the compression time. Let us consider the image matrix $A_{8\times8}$ for explaining the trace of the algorithm. The input parameters for explaining the algorithm are Block size $b=4$, Number of clusters $k=4$, Minimum Support $\psi=2$. The image is split into $n_b=4$ blocks say $B_{A_1}$, $B_{A_2}$, $B_{A_3}$, $B_{A_4}$.
\newline

\begin{center}
\[
A = 
\begin{bmatrix}
139 & 108 & 108 & 108 & 108 & 108 & 108 & 121 \\
177 & 165 & 121 & 139 & 139 & 139 & 121 & 95 \\
239 & 183 & 139 & 139 & 165 & 139 & 108 & 108 \\
225 & 165 & 121 & 165 & 165 & 151 & 108 & 108 \\
165 & 139 & 121 & 152 & 151 & 134 & 121 & 121 \\
113 & 121 & 108 & 121 & 139 & 165 & 183 & 139 \\
108 & 90 & 95 & 134 & 121 & 121 & 165 & 151 \\
95 & 121 & 108 & 108 & 108 & 121 & 139 & 121 
\end{bmatrix}
\]
\resizebox{\linewidth}{!}{%
$
B_{A_1} = 
\begin{bmatrix}
139 & 108 & 108 & 108 \\
177 & 165 & 121 & 139 \\
239 & 183 & 139 & 139 \\
225 & 165 & 121 & 165 
\end{bmatrix}
B_{A_2} = 
\begin{bmatrix}
108 & 108 & 108 & 121 \\
139 & 139 & 121 & 95 \\
165 & 139 & 108 & 108 \\
165 & 151 & 108 & 108 \\
\end{bmatrix}
$
}
\resizebox{\linewidth}{!}{%
$
B_{A_3} = 
\begin{bmatrix}
165 & 139 & 121 & 152 \\
113 & 121 & 108 & 121 \\
108 & 90 & 95 & 134 \\
95 & 121 & 108 & 108 
\end{bmatrix}
B_{A_4} = 
\begin{bmatrix}
151 & 134 & 121 & 121 \\
139 & 165 & 183 & 139 \\
121 & 121 & 165 & 151 \\
108 & 121 & 139 & 121  
\end{bmatrix}
$
}
\end{center}

\begin{table}[!htb]
\centering
 \subfloat[$T_{B_1}$ \label{B1}]{%
      \begin{tabular}{|c|c|}
	\hline
\textbf{Identifier} & \textbf{Mean} \\
\hline
0 & 232 \\
1 & 113 \\
2 & 171 \\
3 & 139 \\
\hline
\end{tabular}
    }
  \subfloat[$T_{B_2}$ \label{B2}]{
  \begin{tabular}{|c|c|}
\hline
\textbf{Identifier} & \textbf{Mean}\\
\hline
0 & 121 \\
1 & 160 \\
2 & 139 \\
3 & 106 \\
\hline
\end{tabular}
\label{B4}
  }
    \\
  
 \subfloat[$T_{B_3}$ \label{B3}]{%
      \begin{tabular}{|c|c|}
	\hline
\textbf{Identifier} & \textbf{Mean} \\
\hline
0 & 123 \\
1 & 93 \\
2 & 109 \\
3 & 152 \\
\hline
\end{tabular}
    }
  \subfloat[$T_{B_4}$ \label{B4}]{
  \begin{tabular}{|c|c|}
\hline
\textbf{Identifier} & \textbf{Mean}\\
\hline
0 & 171 \\
1 & 137 \\
2 & 119 \\
3 & 151 \\
\hline
\end{tabular}
\label{B4}
  }
    \caption{Cluster Identifier Tables}

  \end{table}

\[
A^{'} = 
\begin{bmatrix}
3 & 1 & 1 & 1 & 3 & 3 & 3 & 0 \\
2 & 2 & 1 & 3 & 2 & 2 & 0 & 3 \\
0 & 2 & 3 & 3 & 1 & 2 & 3 & 3 \\
0 & 2 & 1 & 2 & 1 & 1 & 3 & 3\\
3 & 3 & 0 & 3 & 3 & 1 & 2 & 2\\
2 & 0 & 2 & 0 & 1 & 0 & 0 & 1 \\
2 & 1 & 1 & 0 & 2 & 2 & 0 & 3\\
1 & 0 & 2 & 2 & 2 & 2 & 1 & 2 \\ 
\end{bmatrix}
\]

\subsubsection{Image Clustering}
\textit{k}-means clustering aims to partition a given set of data objects into \textit{k} clusters. Every pixel is considered as a data object in the context of an image. The first step in \textit{k}-means is to maximize the intra cluster similarity and the second is to minimize the inter cluster similarity. \textit{k}-means clustering defines \textit{k} centroids, one for each cluster. The centroids should be chosen in such a way that, they are far from each other. Next to this, every data object is associated with the cluster whose centroid is the nearest to the object. After every association, the cluster centroid is updated. Once all the objects are clustered, the global centroids are found and every data object is reassociated to the cluster whose centroid is the nearest, to minimize the inter cluster similarity. This process is repeated until no more reassociations are done. As all the blocks of each component are independent to each other, \textit{k}-means clustering is applied in parallel to all the blocks. The pixels in each block are replaced with their respective cluster identifiers using the block cluster identifier tables.  All the blocks are merged to form the matrix $I^{'}$.  \textit{k}-means is performed on blocks $B_{A_1}, \: B_{A_2}, \: B_{A_3}, \: B_{A_4}$ and the respective block cluster identifier tables are Tables ~\ref{B1}, \ref{B2}, \ref{B3} and \ref{B4}. 

\begin{algorithm}[!htb]
\caption{\textbf{Closed Frequent Sequence Mining}}
\label{CFSM}
\begin{algorithmic}[1]
\State \textbf{Input: }{Sequence Database $I^{'}$, Minimum Support $\psi$}
\State \textbf{Output: }{Closed Frequent Sequences \textit{S}}
\Procedure{\textit{Modified\_GSP}}{}
\State $C_1 \gets \{\sigma|\sigma\:\: is\:\: a\:\: 1-sequence\}$
\State $C_2 \gets C_1 \times C_1$
\State $\forall \gamma \in I^{'}$, for $\sigma \in C_2$, if $\sigma \sqsubseteq \gamma$, \textit{support($\sigma$)}++
\State $F_2 \gets \{\sigma|\sigma \in C_2 \wedge support(\sigma)\geq\psi\}$
\State $l \gets 3$
\While{$F_{l-1}$!=NULL}
\State $C_l$ = $\emptyset$
\For{each ($\sigma \in F_{l-1}$)}
\For{each ($\gamma \in F_{l-1}$)}
\If{($l-2$) length suffix of $\sigma$ == ($l-2$) length prefix of $\gamma$}
\State $\theta\:\: = \:\:\sigma+\gamma[l-1]$ 
\State $C_l = C_l \cup \theta$
\EndIf
\EndFor
\EndFor
\For{each ($\sigma \in C_l)$}
\If{$support(\sigma) \geq \psi$}
\State $F_l \gets F_l \cup \sigma$
\EndIf
\EndFor
\For{each $\gamma \in F_l$}
\For{each $\sigma \in F_{l-1}$}
\If{($\sigma\sqsubseteq\gamma$)$\wedge$($ support(\gamma)==support(\sigma)$)}
\State $F_{l-1} \gets F_{l-1} - \sigma$
\EndIf
\EndFor
\EndFor
\State $l++$;
\EndWhile
\State $S$ = $C_1 \cup F_2 \cup F_3 \cup \dots \cup F_{l-1}$
\EndProcedure
\end{algorithmic}
\end{algorithm}

\subsubsection{Closed Frequent Sequence Mining}
The \textit{support} of a sequence $\sigma$, in a sequence database $D$ is defined as the number of sequences in $D$ that contain $\sigma$,  \textit{support(}$\sigma$\textit{)}=$|\{s|s\in D\:\: \wedge \:\: \sigma \sqsubseteq s\}|$. The sequences whose support is not less than a given support threshold $\psi$ are called as \textit{frequent} sequences \textbf{\textit{FS}}. The set of all \textit{closed frequent} sequences \textbf{\textit{CFS}} is defined as follows, \textbf{\textit{CFS}} = $\{\sigma|\sigma \in FS$ and $\nexists \gamma \in FS$ such that $\sigma \sqsubseteq \gamma$ and $support(\sigma) = support(\gamma)\}$. As the set of closed frequent sequences don't include sequences whose super sequences also have the same support we have $|CFS| \leq |FS|$. Closed Frequent Sequence Mining(CFSM) is to generate all the sequences which are frequent and closed and this can be done in two ways. 1)Mine all the frequent sequences and then prune out all the non closed sequences 2)Generate the closed sequences level wise by pruning the non closed ones\cite{Gomariz2013}. In this paper, method 2 is followed. To achieve this task, Generalized Sequential Pattern(GSP) Mining algorithm for mining frequent sequences is modified to generate closed frequent sequences by executing an additional step at each level to prune out the non closed sequences\cite{srikant1996mining}. In the context of an image, the image matrix is considered to be a sequence database where every row is a sequence of pixels. Support of a sequence $\sigma$ is the number of rows that contain $\sigma$. CFSM is performed on $I^{'}$ to generate the closed frequent sequences. Algorithm~\ref{CFSM} mines closed frequent sequences from $I^{'}$. Table II mines closed frequent sequences from $A$.\\
\par There may be a few characters which may not occur as subsequences in any of the sequences generated for encoding. Those characters may not be encoded leading to an ambiguous encoding. To overcome this issue, sequences in $C_1$ are not pruned. A few sequences and their sub sequences are also frequent. In the example, \textit{3312} and \textit{33} with support 2 and 4 are frequent. As the sequence \textit{3312} includes 2 occurrences of \textit{33}, the support of \textit{33} needs to be modified. If this overlapping support is not modified, the cardinality of the set of closed frequent sequences $S$ may increase in which, many of the sequences which have this overlapping support will not used in the encoding process. The size of the code table will be increased tremendously, leading to a low compression ratio. Consider $H_s$ to be the set of all lengthier sequences and $L_s$ to be the set of proper subsequences from $S$. If the overlapping support of sequences in $L_s$ is not modified, sequences in $L_s$ which actually have lesser support than sequences in $H_s$ will have high support values. Hence the support of these subsequences needs to be modified. Thus the need for finding the modified support($\psi_{mod}$) of every sequence in $S$ is apparent. Sequences in $H_s$ are given more priority than sequences in $L_s$ for encoding, because of which the overlapping support of sequences in $L_s$ is reduced. In order to obtain $\psi_{mod}$, sequences in $S$ are sorted in descending order of their length. For each sequence $\sigma \in S$, find the support and remove all instances of $\sigma$ from $I^{'}$ and update the support of other sequences in $S$ and the process is repeated till $I^{'}$ is empty. The closed frequent sequences with their modified supports form $S^{'}$ and $|S^{'}| \leq |S|$. 
Table III shows the support and modified support of the closed frequent sequences.

\begin{center}
\begin{table*}[h]
\centering
\begin{tabular}{c|c|c}
\hline
\textbf{Length} & \textbf{Candidate Sequence Set} & \textbf{Frequent Sequence Set} \\
\hline
1 & \textbf{$C_1$} = \{0(8), 1(8), 2(7), 3(6)\} & No pruning done\\
\hline 
2 &\textbf{$C_2$} = \{00(1), 01(1), 02(5), 03(3), 10(3), 11(3), & \textbf{$L_2$} = \{02(5), 03(3), 10(3), 11(3), 12(4), 13(3),\\
             &12(4), 13(3), 20(3), 21(4), 22(4), 23(1), &              20(3), 21(4), 22(4), 30(2), 31(3), 33(4)\}\\
             & 30(2), 31(3), 32(1), 33(4)\} & \\
\hline
3 & \textbf{$C_3$} = \{020(1), 021(1), 022(2), 030(0), 031(0),  &\\
          &  033(1), 102(2), 103(0), 110(1), 111(1), 112(0), &\textbf{$L_3$} = \{022(2), 102(2), 113(2), 133(2), 203(2), \\
           &  113(2), 120(0), 121(1), 122(1), 130(0),  131(0), &  211(2), 212(2), 220(2), 221(2), 312(2), 331(2), 330(2)\}\\
            & 133(2), 202(1), 203(2), 210(0), 211(2), 212(2),  &\textbf{$L_2$} = \{02(5), 03(3), 10(3), 11(3), 12(4), 13(3)\}\\
            & 213(1), 220(2), 221(2), 222(1), 310(0), 311(1), &20(3), 21(4), 22(4), 31(3), 33(4)\}\\
            &312(2), 313(0), 330(2), 331(2),333(1)\}  & \\
\hline             
4 & \textbf{$C_4$} = \{0220(1), 0221(0), 1022(2), 1133(2), & \textbf{$L_4$} = \{1022(2), 1133(2), 2203(2), 3312(2)\}\\
         &   1331(0), 1330(0), 2113(1), 2203(2), & \textbf{$L_3$} = \{211(2), 212(2), 221(2), 330(2)\}\\
         &    2211(0), 2212(1), 3312(2)\} & \\
\hline      
\end{tabular}
\caption{CFSM on $A^{'}$ with $\psi=2$}
\end{table*}
\begin{table*}[h]
\centering
\resizebox{\textwidth}{!}{%
\begin{tabular}{|c|c|c|c|c|c|c|c|c|c|c|c|c|c|c|c|c|c|c|c|c|c|c|c|}
\hline
\textbf{Sequence} & 1022 & 1133 & 2203 & 3312 & 211 & 212 & 221 & 330 & 02 & 03 & 10 & 11 & 12 & 13 & 20 & 21 & 22 & 31 & 33 & 0 & 1 & 2 & 3 \\
\hline
\textbf{Support} & 2 & 2 & 2 & 2 & 2 & 2 & 2 & 2 & 5 & 3 & 3 & 3 & 4 & 3 & 3 & 4 & 4 & 3 & 4 & 8 & 8 & 7 & 6  \\
\hline
$\psi_{mod}$ & 2 & 2 & 1 & 2 & 0 & 2 & 1 & 1 & 2 & 1 & 1 & 0 & 0 & 0 & 0 & 1 & 0 & 1 & 1 & 4 & 1 & 3 & 2 \\
\hline
\end{tabular}}
\\
\caption{Closed Frequent Sequences with their Support and Modified Support of $A^{'}$}
\end{table*}
\end{center}
\vspace*{-0.5cm}
\par Consider $E$ to be the matrix where every row is a sequence. Let the closed frequent sequences in $E$ be $\{\textit{333}-2, \textit{11}-2, \textit{44}-2, \textit{1}-2, \textit{2}-1, \textit{3}-2, \textit{4}-2\}$.
\[
E =
\begin{bmatrix}
4 & 4 & 1 & 1 & 2 & 3 & 3 & 3 \\
1 & 1 & 3 & 3 & 3 & 4 & 4 & 3 \\
\end{bmatrix}
\]
As \textit{333} is the sequence with the maximum length, all instances of \textit{333} are removed from $E$ and the support of other sequences is also updated. The closed frequent sequences and their supports are $\{\textit{333}-2, \textit{11}-2, \textit{44}-2, \textit{1}-2, \textit{2}-1, \textit{3}-1, \textit{4}-2\}$. 
\[
E =
\begin{bmatrix}
4 & 4 & 1 & 1 & 2 &  &  &  \\
1 & 1 &  &  &  & 4 & 4 & 3 \\
\end{bmatrix}
\]
The process is repeated until $E$ is empty and the sequences with their modified supports are as follows $\{\textit{333}-2, \textit{11}-2, \textit{44}-2, \textit{1}-0, \textit{2}-1, \textit{3}-1, \textit{4}-0\}$.

\subsubsection{Huffman coding}
Using the sequences in $S^{'}$ with their modified support values, Huffman coding is performed to generate the binary codes for each sequence. The sequences and their corresponding binary codes are stored in Code Table $CT$. The sequences in $S^{'}$ are sorted in descending order of their length. For each sequence $\sigma$ in $S^{'}$, all its instances in $I^{'}$ are replaced with its respective binary code from $CT$ to form the compressed binary image $C_I$. Huffman coding is performed on closed frequent sequences in $A^{'}$. 
The binary encoded image of $A$ is found to be 111 bits in size.
%

\subsection{Decompressor}
Algorithm~\ref{Decompressor} performs decompression. Huffman Decoding is performed on $C_I$ to form $I^{'}$. Every cluster identifier in $I^{'}$ is replaced with the mean from the respective block cluster identifier table to form the decompressed image $D_I$. For $A$, the decompressed matrix is $D_A$.
\begin{algorithm}[!htb]
\caption{\textbf{Decompressor}}\label{Decompressor}
\begin{algorithmic}[1]
\State \textbf{Input: }{
	\begin{enumerate}
		\item Compressed Image $C_I$
		\item Number of blocks $n_b$
		\item Code Table of each component $CT_{1}, CT_{2}, CT_{3}$
		\item Cluster Identifier Table of each block in each component $T_{B_1}, T_{B_2}, \dots, T_B{_{n_b}}$
		
	\end{enumerate}
}
\State \textbf{Output: }{Decompressed Image $D_I$}
\Procedure{\textit{Decompress}}{}
\State Decode each component of the compressed image $C_I$ using the code tables $CT_i$($1\leq i \leq 3$), to get $I_{i}^{'}$
\State For each component, replace each cluster identifier in $I_{i}^{'}$ with the corresponding mean pixel value.
\State Merge all the components to form the decompressed image $D_I$
\EndProcedure
\end{algorithmic}
\end{algorithm}
\begin{minipage}{0.5\textwidth}
\resizebox{\linewidth}{!}{%
$
D_A = 
\begin{bmatrix}
139 & 113 & 113 & 113 & 106 & 106 & 106 & 121 \\
171 & 171 & 232 & 139 & 139 & 139 & 121 & 106 \\
232 & 171 & 139 & 139 & 160 & 139 & 106 & 106 \\
232 & 171 & 113 & 171 & 160 & 160 & 106 & 106 \\
152 & 152 & 123 & 152 & 151 & 137 & 119 & 119 \\
109 & 123 & 109 & 123 & 137 & 171 & 171 & 137 \\
109 & 93 & 93 & 123 & 119 & 119 & 171 & 151 \\
93 & 123 & 109 & 109 & 119 & 119 & 137 & 119\\
\end{bmatrix}
$}
\end{minipage}

\section{Simulation Results and Discussion}
The simulation is performed on standard test images Lena, Peppers, Baboon, Boat, etc. in \textit{bmp} format\cite{imagedataset}. The results are compared with those of JPEG and GIF. The algorithm is implemented in Python 3 and is simulated on 2 cores of Intel core i5-4120 CPU with 1.70GHz clock speed and 4GB Main Memory running Linux Ubuntu 15.10 edition. The results are shown for Lena and Peppers images of size 512$\times$512. The efficiency of compression is affected by the parameters $b$, $k$ and $\psi$ whereas the quality of the decompressed image is affected by $b$ and $k$ only. Hence the effect of $b$, $k$ and $\psi$ are observed on compression efficiency. The compression ratio $C_r$ is defined as 
\begin{center}
$C_{r}$ $ = \frac{\text{\normalsize Uncompressed size of Image}}{\text{\normalsize Compressed size of Image}}$
\end{center}
The image quality is measured using the metrics Peak Signal to Noise Ratio(PSNR), Structural SIMilarity(SSIM) etc. If the average of errors(MSE) is low, PSNR will be high.\\ 


\begin{figure}
	\begin{minipage}{0.24\textwidth}
		\includegraphics[width = \linewidth]{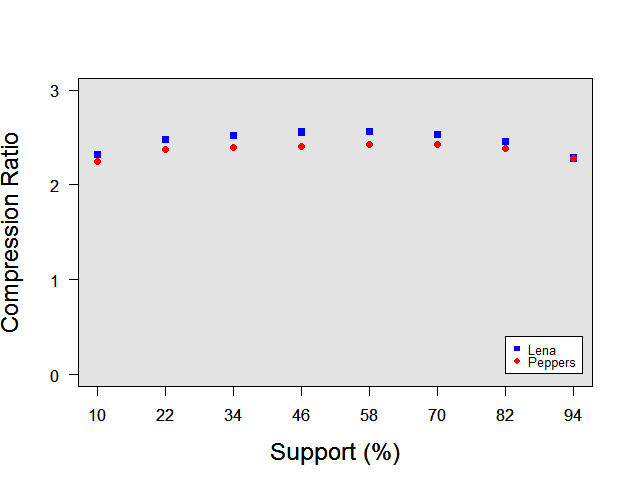}
		\caption{$\psi$ vs $C_r$}
		\label{svscr}
	\end{minipage}
	\begin{minipage}{0.24\textwidth}
		\includegraphics[width = \linewidth]{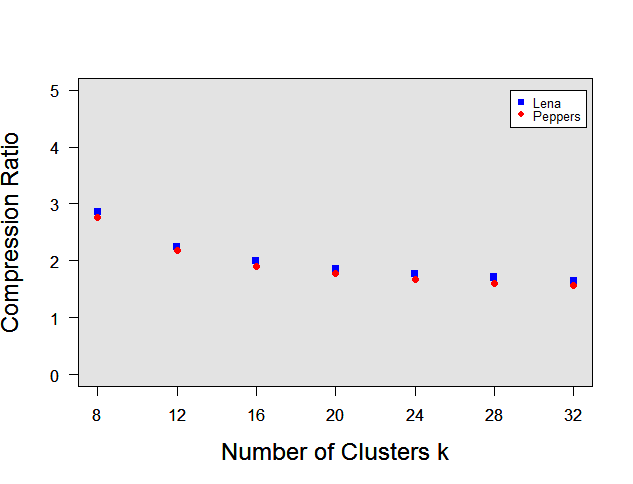}
		\caption{$k$ vs $C_r$}
		\label{kvscr}
	\end{minipage}
	\begin{minipage}{0.24\textwidth}
		\includegraphics[width = \linewidth]{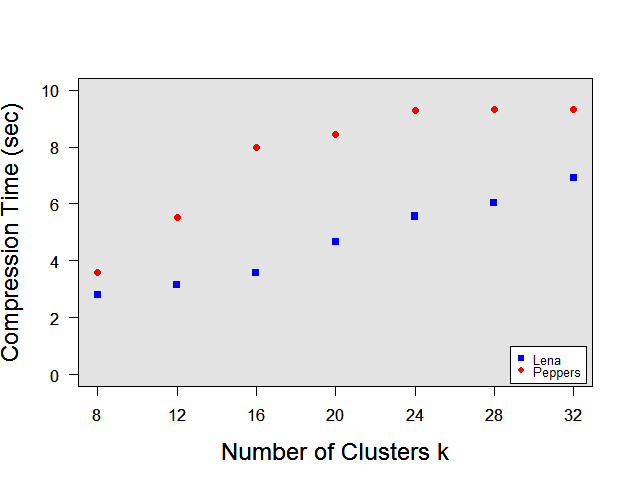}
		\caption{$k$ vs $C_t$}
		\label{kvsct}
	\end{minipage}	
	\begin{minipage}{0.24\textwidth}
		\includegraphics[width = \linewidth]{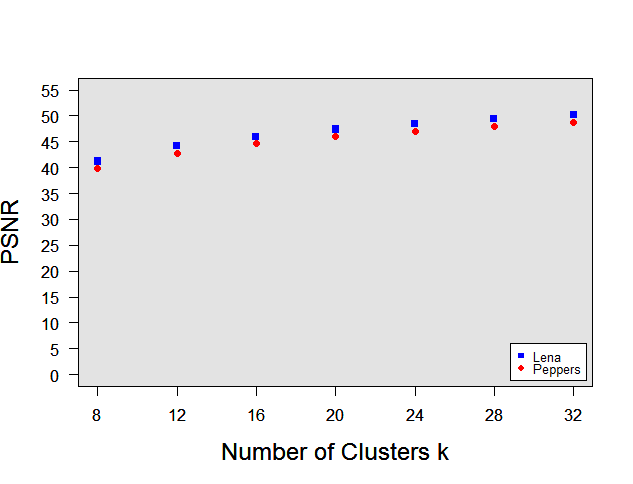}
		\caption{$k$ vs PSNR}
		\label{kvspsnr}
	\end{minipage}
\end{figure} 
\begin{figure}
	
\end{figure}
The compressed size of the image $S_{C_I}$, denotes the overall sum of the sizes of all the Cluster Identifier Tables, Code Table and Encoded Image.
Figure~\ref{svscr} shows the efficiency in terms of Compression Ratio($C_r$) by varying the support($\psi$) for Lena and Peppers. As can be seen, for both Lena and Peppers, the compression ratio increases as the support increases, achieving a maximum and then decreases. Similar trend is observed for other datasets as well. For lower values of $\psi$, $|CFS|$ is more and hence the size of code table is high. For higher values of $\psi$, the $|CFS|$ is less and hence the size of encoded image is high. For Lena, highest compression ratio of $2.555$ is observed for $\psi=$58\%. A compression ratio of $2.425$ is observed at $\psi=70\%$ for Peppers. The compression ratio is observed to be high when $\psi\in[40\%,\:70\%]$. Also, as the block size increases, the compression ratio increases. This is due to a decrease in the number of block cluster identifier tables for an increase in $b$. As the size of each cluster identifier table increases with an increase in $k$, $S_{C_I}$ increases and hence $C_r$ decreases which can be observed from figure~\ref{kvscr}. 
As the value of $k$ increases, time to cluster the pixels increases and hence the compression time which can be observed from figure~\ref{kvsct}. On an average for different values of $b$, the compression time is as low as 2 seconds. This is due to the parallelism brought in clustering blocks. Figure ~\ref{kvspsnr} show the effect of $k$ on PSNR for Lena and Peppers. As the number of clusters increases, the quality of the image increases due to grouping of pixels into more closely related clusters. For $k=32$, PSNR is $50.26$ for Lena and $48.70$ for Peppers. It is also observed that as block size increases PSNR decreases. 

%

\begin{figure}[h]
\begin{center}
%
  \subfloat[Proposed \label{224kB1}]{%
      \includegraphics[width=0.16\textwidth]{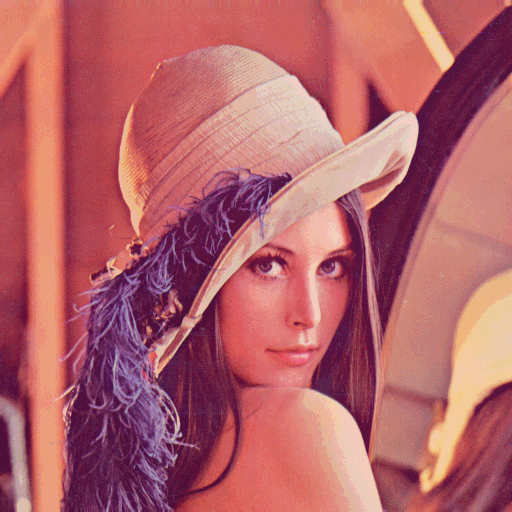}
    }
    \hspace*{0.5cm}
 \subfloat[JPEG \label{224kB2}]{%
      \includegraphics[width=0.16\textwidth]{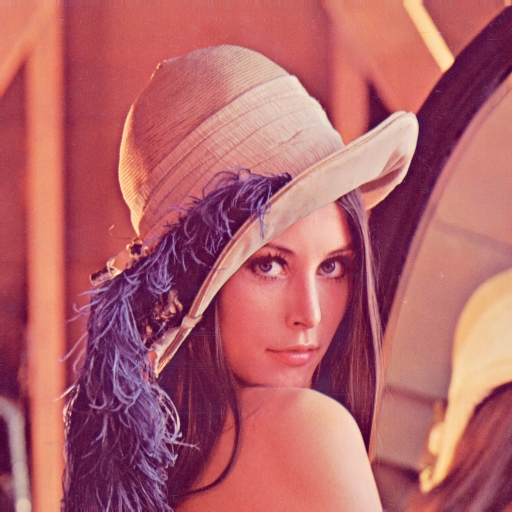}
    }
    \caption{Lena compressed Images at 224kB}
\end{center}
\end{figure}
Tables IV and V show the comparison of the proposed algorithm with JPEG and GIF algorithms for Lena and Peppers images respectively. The size of the input image is 786.6kB. For Lena, at size 224kB, the PSNR of the proposed algorithm is 39.49 whereas that of JPEG is 41.82. The quality of the decompressed image is found to be close to that of JPEG at 224kB. The images are shown in figures~\ref{224kB1} and \ref{224kB2}.
\FloatBarrier
\begin{table*}[!htbp]
\centering
\resizebox{\textwidth}{!}{%
\begin{tabular}{c |c |c |c |c |c |c |c |c}
\hline
\textbf{Image} & \textbf{Input Size}  & \textbf{Proposed Size} & \textbf{JPEG Size} &\textbf{PSNR-Proposed} & \textbf{PSNR-JPEG} & \textbf{SSIM-Proposed} & \textbf{SSIM-JPEG} & ($\boldsymbol{b,k,\psi}$)\\
 & \textbf{(kB)} &\textbf{(kB)} & \textbf{(kB)} &\textbf{ No Unit} & \textbf{No Unit} & \textbf{No Unit} & \textbf{No Unit} & \\
\hline
\hline
Lena & 786.6 & 166.7 & 162 & 37.72 & 41.19 & 0.9697 & 0.9870 & (128,8,70) \\
 	 & 786.6 & 224 & 224 & 39.49 & 41.82 & 0.978 & 0.992 & (256,12,70) \\
 	 & 786.6 & 359 & 341.3 & 45.99 & 47.92 & 0.997 & 0.995 & (32, 16, 82) \\
Peppers & 786.6 &  241 & 242 & 38.09 & 38.45 & 0.973 & 0.989 & (256,12,82) \\
        & 786.6 & 254 & 262 & 38.99 & 39.19 & 0.982 & 0.99 & (128,12,70) \\
        & 786.6 & 365 & 363 & 44.58 & 46.27 & 0.996 & 0.996 & (32,16,94) \\
\hline
\end{tabular}}
\label{JPEGComparision}
\caption{Comparison of Proposed Algorithm with JPEG for Lena and Peppers}
\end{table*}
\begin{table*}
\centering
\resizebox{\textwidth}{!}{%
\begin{tabular}{c |c |c |c |c |c |c |c |c}
\hline
\textbf{Image} & \textbf{Input Size}  & \textbf{Proposed Size} & \textbf{GIF Size} &\textbf{PSNR-Proposed} & \textbf{PSNR-GIF} & \textbf{SSIM-Proposed} & \textbf{SSIM-GIF} & ($\boldsymbol{b,k,\psi}$)\\
 & \textbf{(kB)} &\textbf{(kB)} & \textbf{(kB)} &\textbf{ No Unit} & \textbf{No Unit} & \textbf{No Unit} & \textbf{No Unit} & \\
\hline
\hline
Lena & 786.6 & 221.3 & 226 & 40.64 & 29.55 & 0.985 & 0.78 & (128, 12, 58) \\
     & 786.6 & 205 & 204 & 39.49 & 28.78 & 0.978 & 0.77 & (256, 12, 46) \\
     & 786.6 & 173.45 & 173.71 & 37.72 & 27.12 & 0.969 & 0.75 & (128, 8, 46) \\
Peppers & 786.6 & 220.13 & 216 & 38.096 & 28.94 & 0.973 & 0.81 & (256, 12, 46) \\
		& 786.6 & 207.59 & 204 & 35.14 & 28.13 & 0.94 & 0.79 & (256, 8, 94) \\
		& 786.6 & 185.38 & 192 & 36.088 & 27.94 & 0.95 & 0.78 & (128, 8, 46) \\
\hline
\end{tabular}}
\label{GIFComparision}
\caption{Comparison of Proposed Algorithm with GIF for Lena and Peppers}
\end{table*}
For Peppers, at the compressed image size 241kB, the PSNR of the proposed algorithm and that of JPEG have a negligible difference and the SSIM values are also close. Images with quality close to that of JPEG are produced by the proposed algorithm for the tested datasets. For Lena, the compressed image of size 221.3kB has a PSNR of 40.64 and an SSIM of 0.985 by the proposed algorithm whereas the compressed image in GIF format of size 229kB has a PSNR of 29.55 and SSIM of 0.78. For Peppers, the compressed image of size 220.13kB has a PSNR of 38.09 and an SSIM of 0.973 by the proposed algorithm whereas the compressed image in GIF format of size 204kB has a PSNR of 28.13 and SSIM of 0.81. The proposed algorithm always outperforms GIF in terms of compression ratio and quality. Other datasets also depicted the similar trend. 
%
%
%
%

%

\section{Conclusion}
An efficient and novel frequent sequence mining based lossy image compression algorithm is presented in this paper. Simulation results have proved that our work is better in mining optimal pixel sequences outperforming some of its alternatives in compression ratio with an acceptable loss in visual quality. As a part of future work, other efficient clustering algorithms can be tried in place of \textit{k}-means to improve the visual quality of the image. Neighbourhood properties of the image can be exploited to improve the compression ratio. We aim to conduct extensive simulations on standard datasets with large size to investigate several other parameters including the compression time. 

\bibliographystyle{plain}
\bibliography{references}

\end{document}